\documentclass[prd,aps,floats]{revtex4}

\usepackage{amsmath}

\usepackage{amssymb}

\numberwithin{equation}{section}

\usepackage{epsfig}

\begin{document}

\def\beq{\begin{equation}}

\def\eeq{\end{equation}}

\def\ber{\begin{eqnarray}}

\def\eer{\end{eqnarray}}

\def\l{\Lambda}

\def\b{{\rm b}}

\def\m{{\rm m}}

\def\om{\Omega_{0\rm m}}

\def\oml{\Omega_l}

\def\omx{\Omega_{\l_{\rm b}}}

\def\eg{8 \pi G}

\def\ppp{{\prime\prime\prime}}

\def\pppp{{\prime\prime\prime\prime}}

\def \lleq {\lower0.9ex\hbox{ $\buildrel < \over \sim$} ~}

\def \ggeq {\lower0.9ex\hbox{ $\buildrel > \over \sim$} ~}

\def\rhoc{\rho_{0 {\rm c}}}

\def\t{\times}

\def\etal{{\it et al.}}

\def\ie{i.e.~}

\def\n{\noindent}

\def\apj{Astroph.~J.~}
\def\mn{Mon.~Not.~Roy.~Ast.~Soc.~}
\def\asta{Astron.~Astrophys.~}
\def\aj{Astron.~J.~}
\def\prl{Phys.~Rev.~Lett.~}
\def\prd{Phys.~Rev.~D~}
\def\nucp{Nucl.~Phys.~}
\def\nat{Nature~}
\def\plb{Phys.~Lett.~B~}
\def\jetpl{JETP ~Lett.~}

\title{Is a step in the primordial spectral index favoured by CMB data ?}

\author{Minu Joy$^{a,b}$, Arman Shafieloo$^{a,c}$, Varun Sahni$^a$ and
Alexei A. Starobinsky$^{d}$}

\affiliation{$^a$
Inter-University Centre for Astronomy and Astrophysics,
Pune 411~007, India}
\affiliation{$^b$
Department of Physics, Alphonsa College, Pala, Kerala}
\affiliation{$^c$
Department of Physics, University of Oxford, 1 Keble Road, Oxford, OX1 3RH, UK
}
\affiliation{$^d$
Landau Institute for Theoretical Physics, 119334 Moscow,
Russia }

\thispagestyle{empty}

%\maketitle

\sloppy

\begin{abstract}
\small{
A sudden small change in the second derivative of the inflaton potential
can result in a universal local feature in the spectrum of primordial
perturbations generated during inflation. The exact solution describing
this feature \cite{minu} is characterized by a step in the spectral index
modulated by characteristic oscillations and results in a large running of the
spectral index localized over a few e-folds of scale. In this paper we
confront this step-like feature with the 5 year WMAP results and demonstrate
that it provides a better fit to this data than a featureless initial spectrum.
If such a feature exists at all, then it should lie at sufficiently large scales
$k_0 \lesssim 0.003 {\rm Mpc}^{-1}$ corresponding to $l \lesssim 40$. The sign
of the effect is shown to correspond to the negative running of $n_s$
localized near this scale. This feature could arise as a result of a
`mini-waterfall'-type fast second order phase transition
experienced by an auxiliary heavy field during inflation, in a model
similar to hybrid inflation (though for a different choice of parameters).
If this is the case, then the
auxiliary field should be positively coupled to the inflaton.
}

\end{abstract}

\maketitle

\section{Introduction}
\label{sec:intro}
The great precision of current cosmological data and the enormous volume
of data expected in coming years leads to the hope that cosmological
parameters will soon be determined to great accuracy. Since cosmological
parameters are intimately linked to an underlying theoretical model, the
increasing depth and sophistication of cosmological data sets, especially
those associated with the Cosmic Microwave Background (CMB), will, in all
likelihood, lead to a more refined and deeper understanding of such
important issues as the form of the initial perturbation spectrum, the energy
scale of inflation, etc.

Indeed, the cosmological world view which has emerged during the past two
decades has been quite intriguing! New cosmological data has, while
providing support to earlier constructs such as inflation, also paved the way
for radically new propositions such as dark energy.

%WMAP3$^1$

%{\footnotetext[1]{Third year data release of Wilkinson Microwave Anisotropy

%Probe (WMAP)}}

An important issue concerns the form of the primordial spectral index
$n_s(k) \equiv d\ln P(k)/d\ln k$. The large quantity of CMB data which has become
available over the past few years indicates that the departure of the
spectral index from exact scale invariance is likely to be small,
$|n_s(k)-1|\ll 1$, which is in good agreement with predictions of the simplest
inflationary scenarios (see e.g. \cite{LSV07} for a recent model-independent
analysis of observational data assuming local analytic behaviour of the
inflaton potential $V({\phi})$ in the observable window).  An associated
question concerns the value of the running of the spectral index
$\alpha(k)\equiv d\,n_s(k)/ d\ln k$. Most inflationary models with smooth
inflaton potentials do predict a non-vanishing value of $\alpha(k)$, usually
$|\alpha(k)|\sim |n_s(k)-1|^2\ll 1$. However, recent WMAP$^1$ {\footnotetext[1]
{Wilkinson Microwave Anisotropy Probe (WMAP)}} results
\cite{wmap,wmap5,komatsu} may be suggesting a somewhat larger value
$|\alpha(k)|\sim |n_s(k)-1|$. Taken together with other `anomalies' such as
the `Archeops feature' at $l\sim 40$ \cite{archeops,wmap,wmap5},
these recent data may be providing a subtle hint that inflationary models are
slightly more complex than the simplest single-field models suggested during
the early 1980's.

Note that the only way for a large value of the running $\alpha(k)$ to be
accommodated within the inflationary paradigm is for it to be localized (\ie
have the form of a {\em feature}), since, otherwise, the number of
inflationary e-foldings turn out to be too small \cite{EP06}. Generation of
features in the primordial inflationary spectrum has been the subject of
considerable discussion. A partial list of references may be found in
\cite{feature,S92,sarkar,G05,YY06}. Several model independent attempts to
search for such features in the primordial spectrum directly from the WMAP
data have led to some positive, though inconclusive, results \cite{directpps}.
In this paper we focus on the exact solution for the primordial scalar
perturbation spectrum generated during inflation found in \cite{minu}, in which
the effective mass of the inflaton, $V''({\phi})$, experiences a sudden small
change. This model assumes a much smoother potential than
those which had previously been
considered, which dealt either with a sudden step in the potential \cite{S92,sarkar,
G05}, or in its first derivative \cite{S92,G05}. As a result, the corresponding
feature in the perturbation spectrum is the weakest (in particular,
superimposed oscillations are very small). Still it is observable, and can even
produce the dominant contribution to the running of the slope of the
power spectrum
over the observable range of scales. We shall show that this new universal
(i.e. not dependent on details of the change) feature in the inflationary
perturbation spectrum is in excellent agreement with recent WMAP5
results.

A microscopic model producing such feature in the inflaton potential may be
chosen similar to that used in the hybrid inflationary scenario \cite{hybrid},
but with different numerical values of parameters in order that a fast
(quasi-equilibrium) second order phase transition experienced by another heavy
scalar field should occur {\em during} inflation, and not at its end. Also
this second field should be sufficiently weakly coupled to the inflaton to
produce a small change, $|\Delta m| \ll H$, in the inflaton mass where $H$ is
the Hubble parameter, $H\equiv \dot a/a$, at the moment of the transition. Thus, such a
transition may be called a `mini-waterfall' in analogy with the  `waterfall'
transition used to end inflation in the hybrid inflationary scenario. The
sign of this coupling may be arbitrary, in principle, but we shall show that
the (more natural) positive coupling is favoured by the WMAP5 data since it
corresponds to the negative running of the power spectrum.

\section{Inflation with mini-waterfall is tested using CMB data }
\label{sec:test}

Let us first remind the reader of the microscopic two-field model used in
\cite{minu} to produce a step in the second derivative of an effective inflaton potential.
This is just the well-known model with a Higgs-like potential, also used in the
hybrid inflationary scenario:
\beq V(\psi,\phi) = \frac{1}{4\lambda}\left (M^2 -
\lambda\psi^2\right )^2 + \frac{1}{2}m^2\phi^2 +
\frac{g^2}{2}\phi^2\psi^2~, \label{eq:hybrid} \eeq
(as pointed above, a positive value of $g^2$ is favoured by the data, so
the notation $g^2$ for the coupling parameter is justified). Close to $\psi = 0$,
the effective mass of the field $\psi$ is
\beq
m_\psi^2 \equiv \frac{d^2V}{d\psi^2} = g^2\phi^2 -M^2~.
\eeq
At $\phi_c = M/g$ the curvature of $V(\psi,\phi)$ along the $\psi$ direction
vanishes so that $m_\psi^2  > 0$ for $\phi > \phi_c$ while $m_\psi^2  < 0$
for $\phi < \phi_c$. This implies that for large values of the inflaton
$\phi$ the auxiliary field $\psi$ rolls towards $\psi = 0$.
However, once the value of $\phi$ falls below $\phi_c$ the $\psi = 0$,
configuration is destabilized resulting in a rapid cascade (mini-waterfall)
which takes $\psi$ from $\psi = 0$ to its zero-temperature equilibrium value
$\psi^2=(M^2- g^2\phi^2)/\lambda$ where its mass becomes positive and large
once more. We assume that the field $\psi$ is sufficiently heavy, so it
quickly relaxes to its equilibrium value both before and after the
mini-waterfall during a characteristic time $\delta t\ll H^{-1}$. For this to
occur, the parameters of the model (\ref{eq:hybrid}) should satisfy a number
of inequalities which were presented and investigated in \cite{minu}. In
particular, $g^2$ should be much less than $\lambda$.

If so, quantum fluctuations of the field $\psi$ may be neglected. Then $\psi$
may be excluded from the action of the model (\ref{eq:hybrid}), and its
non-zero equilibrium value $\psi(\phi)$ after the transition leads to the
change in the effective inflaton potential: $V(\phi,\psi)\to V(\phi,\psi(\phi))
\equiv \tilde V(\phi)$ (we omit tilde in what follows). Similar to free energy
during a second order phase transition in thermodynamic equilibrium, the
potential $V$ and its first derivative with respect to $\phi$ are continuous
at $\phi=\phi_c$, while its second derivative acquires a step at this point.
This is, in particular, how it is done in the theory
of hybrid inflation \cite{hybrid} (see also \cite{YY06} for a more general
case). However, in contrast to hybrid inflation, the change in the effective
mass square is small as compared to $H^2$, so inflation continues without a
break in our scenario.

\begin{figure}[tbh!]
\centerline{ \psfig{figure=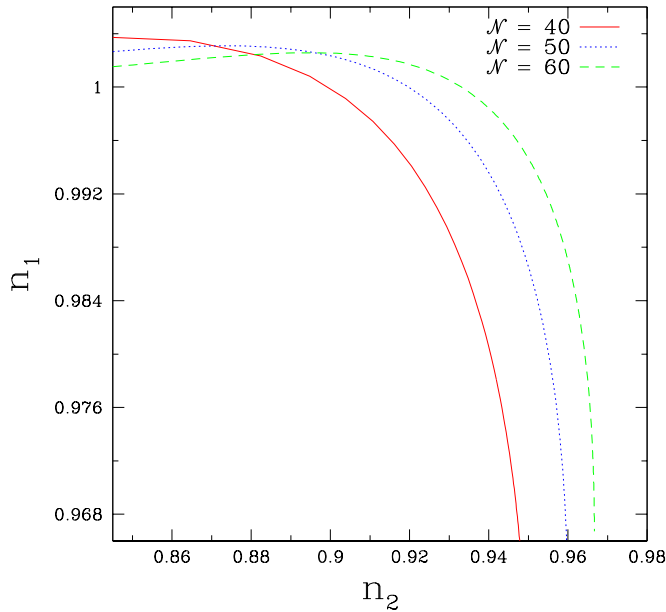,width=0.5\textwidth,angle= 0} }
\caption{\label{fig:n1n2}
Spectral indices for perturbations generated just before ($n_1$) and
immediately after ($n_2$) the phase transition are shown for three values of
the number of inflationary e-folds occurring during the post-mini-waterfall
period:
${\cal N} = 40$ (red, solid),
${\cal N} = 50$ (blue, dotted) and ${\cal N} = 60$ (green, dashed).
}
\end{figure}

In this case, as shown in  \cite{minu}, the values of the primordial spectral
indices $n_1$ and $n_2$ of the power spectrum of scalar perturbations induced
by quantum fluctuations occurring before and after the mini-waterfall are
given by
\ber
n_1 - 1 &=& \frac{1}{2\pi}\left (\frac{g m_P}{M}\right
)^2\frac{\kappa~(1-2\kappa)}
{(1+\kappa)^2} ~,\nonumber\\
n_2-1 &=& -\frac{1}{2\pi}\left (\frac{g m_P}{M}\right)^2
\frac{4+3\kappa+2\kappa^2}{(1+\kappa)^2} ~, \nonumber\\
n_1-n_2 &=& \frac{2}{\pi (1+\kappa)}\left(\frac{gm_P}{M}\right)^2>0~.
\label{eq:n2}
\eer
The parameter combinations $\kappa \equiv 2\lambda m^2/g^2M^2$ and $g m_P/M$
which appear in (\ref{eq:n2}) also enter into the expression for the number
of inflationary e-folds which take place during the post-mini-waterfall period
\beq
{\cal N} = \pi \left ( \frac{M}{gm_P}\right )^2\left [1 + \left
(1+\frac{\kappa}{2}\right ) \log \frac{2+\kappa}{\kappa}\right]~.
\label{eq:N}
\eeq
In fig. \ref{fig:n1n2}, the pair $\lbrace n_1,n_2\rbrace$ is shown for
different values of ${\cal N}$.

Having demonstrated that a step in the value of $n_s$ can be accommodated
within the model (\ref{eq:hybrid}), one still needs to work out the precise
form of the power spectrum $P_R(k) \propto k^{n_s-1}$. This was accomplished
in \cite{minu} where the perturbation equation for gauge invariant quantities
was solved exactly. We summarize some of these results below.

\begin{figure}[tbh!]
\centerline{\psfig{figure=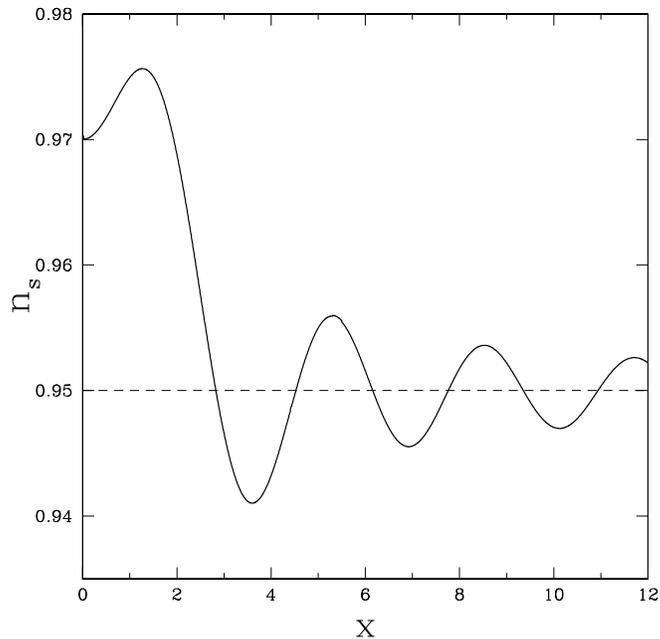,width=0.5\textwidth,angle=0} }
\caption{\small
A step in the second derivative of the inflaton potential leads to a step
in the spectral index as shown in this figure which plots the primordial
spectral index $n_s$ as a function of $x = k/k_0$. The step in $n_s$ at
$x \sim 1$ is followed by oscillations with decreasing amplitude. The
parameters shown here correspond to $n_1 = 0.97$, $n_2 = 0.95$,
which agree well with WMAP5 data.
}
\label{fig:ns}
\end{figure}

In the absence of a phase transition, the power spectrum of cosmological
perturbations generated during inflation has the form $\mathcal{P_R}_0(k)
\propto k^{n_2-1}$. As shown in \cite{minu}, the effect of the phase
transition is to alter $\mathcal{P_R}_0(k)$ by a factor $|\alpha-\beta|^2$,
so that the final spectrum becomes
\beq
P_R(k) \propto \mathcal{P_R}_0(k)\times |\alpha-\beta|^2
\label{eq:power_spectrum}
\eeq
where $n_2$ is the value of the spectral index at $k \gg k_0$, $k_0$ being
the location of the feature. The Bogoliubov coefficients $\alpha, \beta$ can
be expressed in closed analytical form as follows \cite{minu}
\ber
\alpha-\beta &=& -\frac{i\pi \Delta}{2}H_{\mu_1}^{(2)}(k\eta_0)J_{\mu_2}(k\eta_0)
-\frac{i\pi k\eta_0}{2}\left \lbrack H_{\mu_1+1}^{(2)}(k\eta_0)J_{\mu_2}(k\eta_0)
-H_{\mu_1}^{(2)}(k\eta_0)J_{\mu_2+1}(k\eta_0)\right\rbrack~,\\
\nonumber\\
\alpha+\beta &=& \frac{\pi \Delta}{2}H_{\mu_1}^{(2)}(k\eta_0)Y_{\mu_2}(k\eta_0)
+ \frac{\pi k\eta_0}{2}\left\lbrack H_{\mu_1+1}^{(2)}(k\eta_0)Y_{\mu_2}(k\eta_0)
-H_{\mu_1}^{(2)}(k\eta_0)Y_{\mu_2+1}(k\eta_0)\right\rbrack~,\\
\nonumber\\
|\alpha|^2 - &|\beta|^2& = 1~, \label{eq:bog2}
\eer
where $\Delta=\mu_2-\mu_1 =(n_1-n_2)/2$,
~$\mu_{1,2} = \frac{3}{2} - \frac{V''_\mp}{3H_0^2} + 3 \epsilon_0 $,
and $\mp$ denotes the value of $V'' \equiv \frac{d^2V}{d\varphi^2}$
at $t=t_0\mp 0$, $t_0$ being the time of the phase transition. The resulting form of
the spectral index in the vicinity of the feature
has a step-like discontinuity modulated by small oscillations, as shown
in figure \ref{fig:ns}.

%\clearpage

\section{Results}
\begin{figure}[tbh!]
\centerline{\psfig{figure=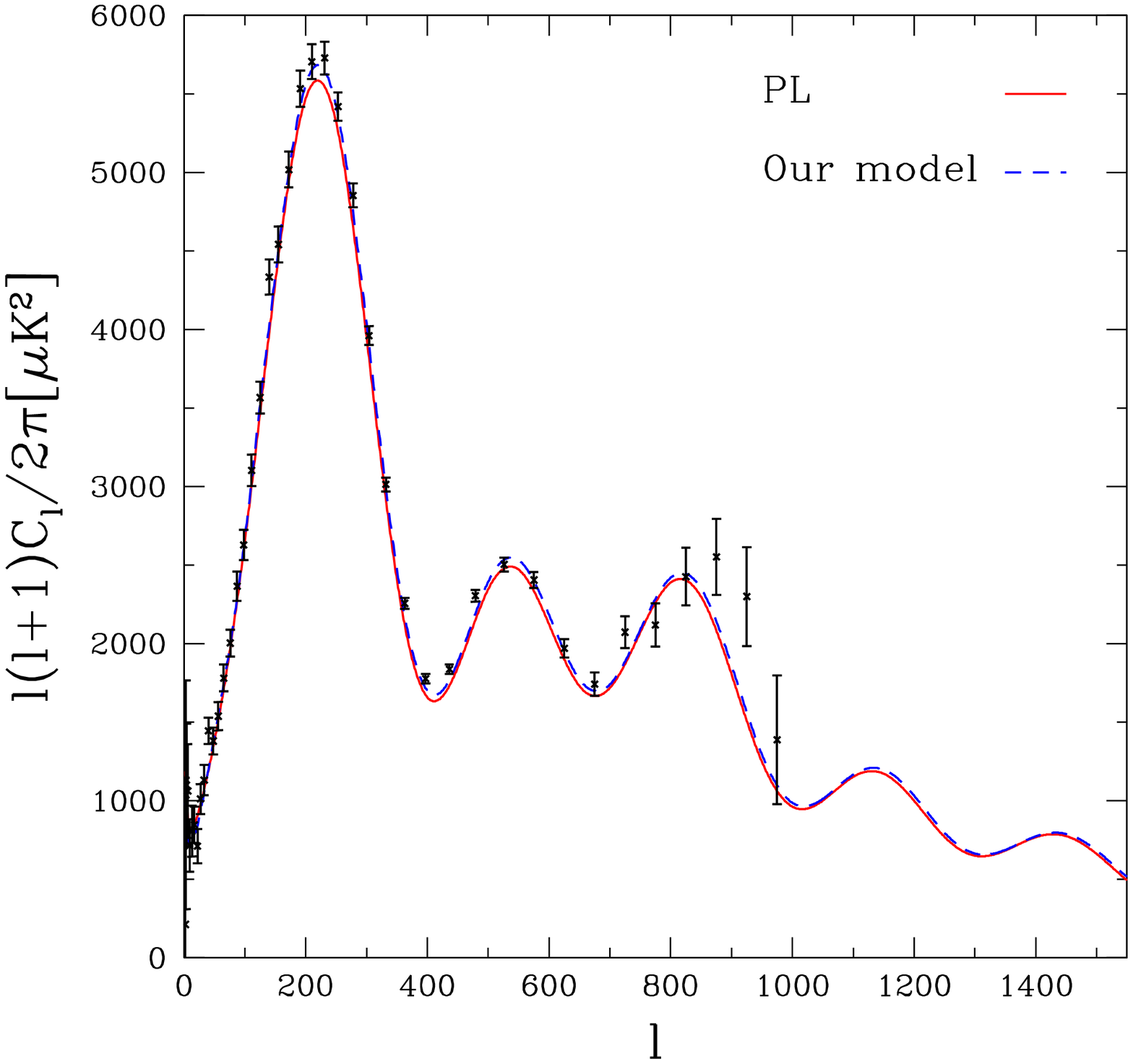,width=0.5\textwidth,angle=0}
\psfig{figure=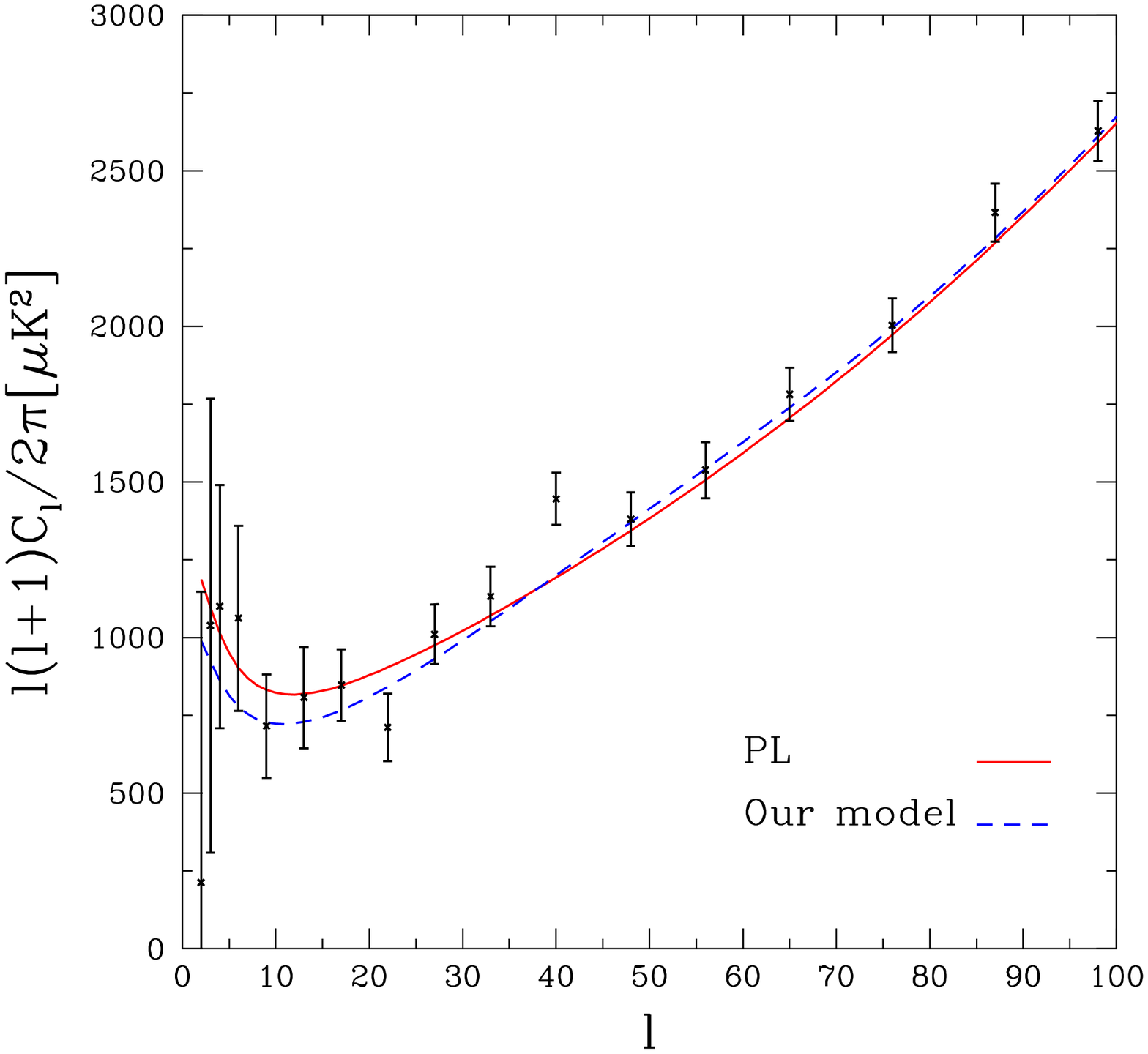,width=0.5\textwidth,angle=0} }
\caption{\small
Comparison of our model~(blue, dashed) with a pure power-law model~(red, solid),
for the best fit values of parameters. The WMAP5 binned
data with related error bars are also plotted for comparison.}
\label{fig:Cls}
\end{figure}

\begin{figure}[tbh!]
\centerline{\psfig{figure=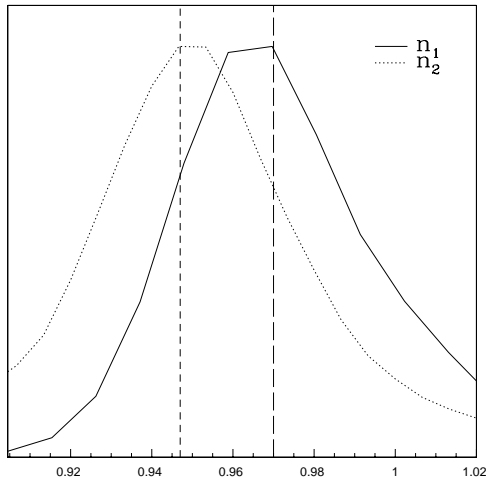,width=0.4\textwidth,angle= 0}
}
\caption{\small
Marginalized posterior distributions for the spectral indices $n_1$ and $n_2$.
}
\label{fig:ns1}
\end{figure}

Fig. \ref{fig:Cls} compares our model (local running) with a Power Law
model~(PL). Results shown are for the best fit values of the cosmological parameters
determined assuming flat $\Lambda$CDM as the background metric.
The package CosmoMC \cite{cosmomc} was used
to confront our model with the five-year observational data from WMAP. We find
that our model gives a better fit with $\Delta \chi^2_{eff} = -3.052$ as
compared to the best PL model with a constant $n_s$. This improvement in
the value of the likelihood has been obtained by introducing two more
free parameters in our model as compared to the PL model, therefore it is
reasonable. In fig. \ref{fig:ns1}, the solid lines are marginalized
probabilities of $n_1$ (the spectral index before the phase transition)  and
the dotted line gives the marginalized probabilities of $n_2$
(the spectral index after the phase transition). Observational data
for multipoles with $l\ge 2$ have been used. A
contribution from primordial tensors, satisfying the one-inflaton
consistency relation, $n_t = -r/8$, has been taken into account, too. The
pivot point is set at $0.05Mpc^{-1}$.

\begin{figure}[tbh!]
\centerline{\psfig{figure=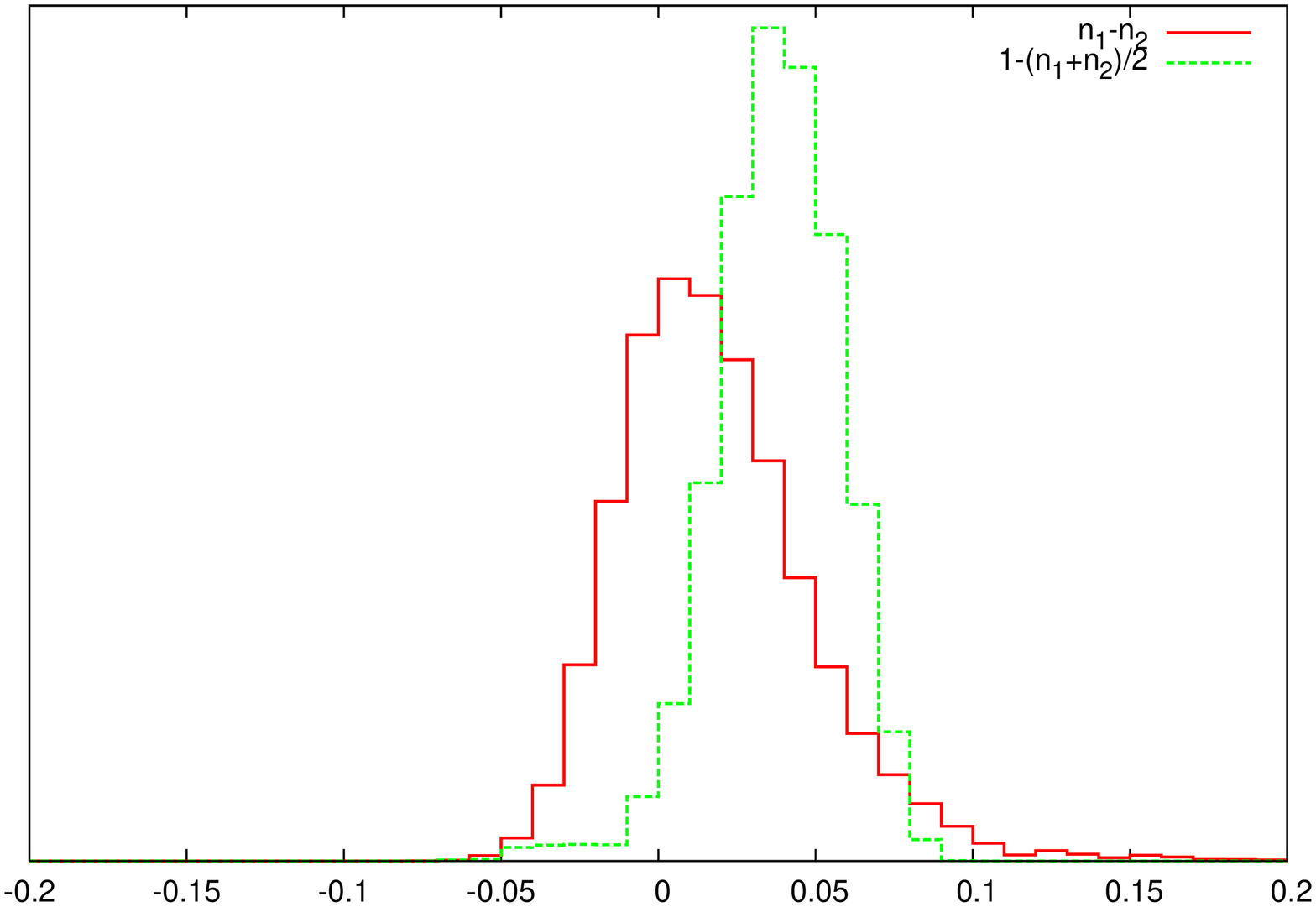,width=0.3\textwidth,angle=-90}
\psfig{figure=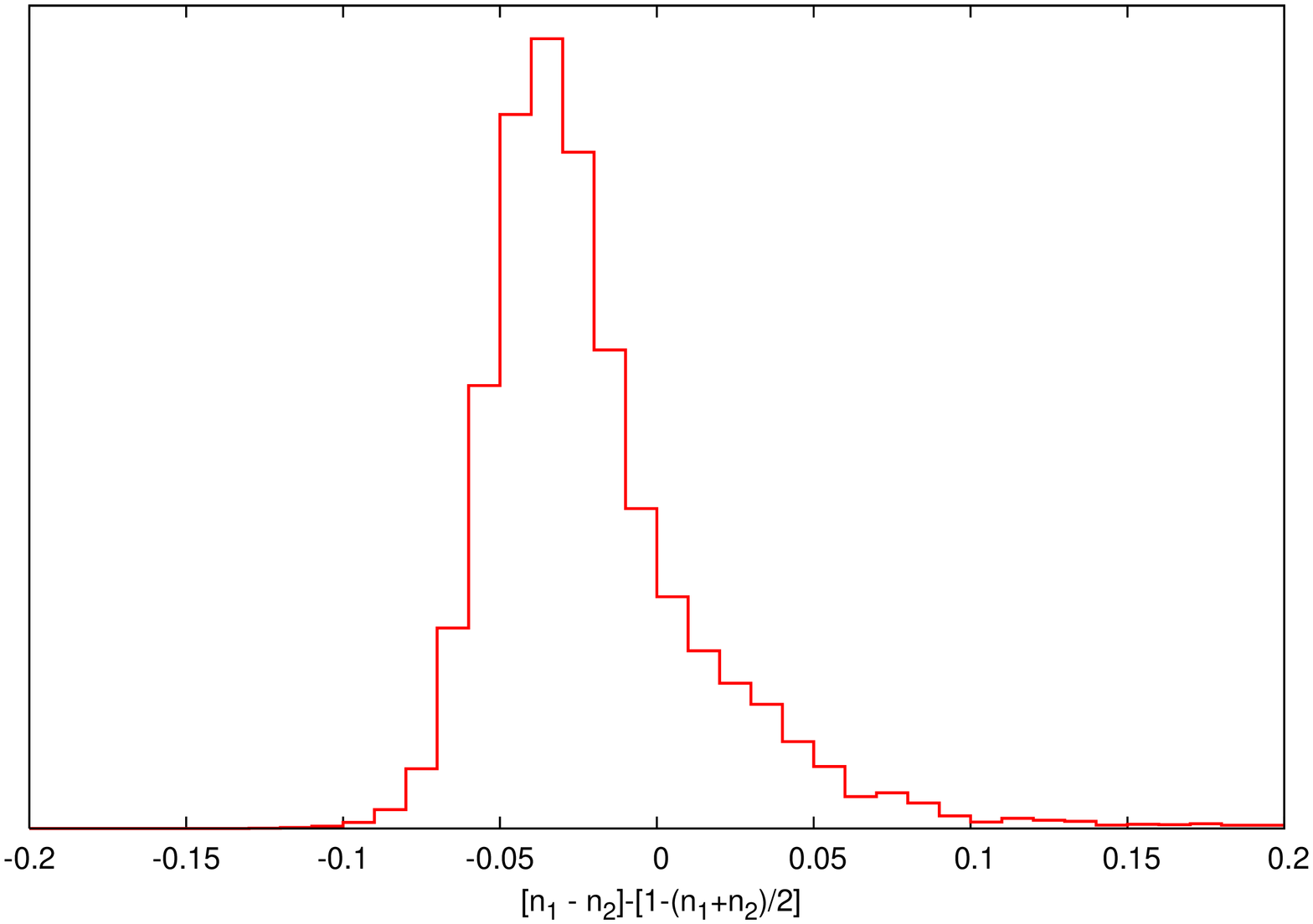,width=0.3\textwidth,angle=-90} }
\caption{\small Marginalized posterior distributions for $(n_1 -
n_2)$ and also for $(1-(n_1+n_2)/2)$. The right panel gives the
marginalized distribution of $[(n_1-n_2) - (1-(n_1+n_2)/2)]$. }
\label{fig:n1n2M}
\end{figure}

 From fig. \ref{fig:ns1} it is clear that a value of $n_1$ slightly larger than that
without a step is favoured by the data. The marginalised probability for $n_2$ peaks
at 0.947, while the marginalized probability for $n_1$ peaks at 0.97. Looking at
the marginalized probability itself, we can see why a local running is suggested
by our model: the peak of $n_1$ lies at a value significantly larger
than $n_2$. It is also interesting that the $\{n_1, n_2\}$ pair,
favoured by CMB data, can be accommodated by the model (\ref{eq:hybrid}) for
very reasonable values of ${\cal N}$ lying in the interval 40 - 60, as shown
in fig. \ref{fig:n1n2}. In particular, to get just the peak values for $n_1$
and $n_2$ given above, one has to choose $gm_P/M\approx 0.42$ and $\kappa
\approx 3.8$ that leads to ${\cal N}\approx 40$.
The distributions for $(n_1 - n_2)$ (red solid line) and also for $(1-(n_1+n_2)/2)$
(green dashed line) marginalized over all the other parameters are given in the left
panel of fig. \ref{fig:n1n2M}. It is quite clear from the plot that the data prefers
$(n_1 - n_2)$ be smaller than $(1-(n_1+n_2)/2)$. Interestingly, the case
$(n_1-n_2) < (1- (n_1+n_2)/2)$ corresponds to $\kappa > 2$ in our model. That is, the
potential energy of the massive inflaton $m^2\phi^2/2$ dominates the vacuum energy
density $V(0,0)=M^4/4\lambda$ at the moment of the phase transition. In the right panel
of fig. \ref{fig:n1n2M} one can see the distribution of $[(n_1-n_2) - (1-(n_1+n_2)/2)]$
marginalized over all the other parameters. It is clear that the peak is at a value less
than zero confirming the above result.

Note that in our previous theoretical paper \cite{minu}, we were
more concerned with the case $\kappa \le 1$ since it leads to a
larger value of $n_1-n_2$, in particular, the spectrum even
becomes blue-tilted for $k\to 0$ if $\kappa<1/2$. But
observational data appear to have made their verdict for the
opposite case $\kappa > 1$ when the potential $V(\psi,\phi)$ is
dominated by the $m^2\phi^2/2$ term during the phase transition.

However, all formulae in \cite{minu} and in Sec. 2 are valid for
any value of $\kappa$ provided it is not very large. In more
detail, the assumption of a fast phase transition is valid if
$m_{\psi}^2\gg H^2$ during the transition, which requires
$1+\kappa \ll \lambda mm_P^2M^{-3}$ (Eq. (4.9) of the paper
\cite{minu}). For the typical values $\lambda\sim 0.1,~m\sim
10^{-6}m_P,~M\sim 10^{-3}m_P$ (see the table 1 in \cite{minu}),
this leads to $\kappa \ll 100$ and $g^2 \ll 10^{-5}$, while
$\kappa=1$ corresponds to $g^2=2\times 10^{-7}$. It can be
verified that if this condition is satisfied, then the
contribution of $\psi$ particles created during this transition to
the total energy density is small compared to the change in the
equilibrium value of $V(\psi(t), \phi(t))$ due to the step in its
second derivative for the time period $\delta t \sim 1/H$ after
the transition during which the correction
(\ref{eq:power_spectrum}-\ref{eq:bog2}) to the quasi-flat power
spectrum $\mathcal{P_R}_0(k)$ is generated, irrespective of
whether these created $\psi$ particles have decayed into other
particles over this time period or not.

Another inequality bounding $\kappa$ from above follows from the
requirement that the change in the spectral index $n_1-n_2$ should
be much more than next-order slow-roll corrections to $n_1$ and
$n_2$ (not taken into account in our calculations), which are
$\sim {\cal N}^{-2}$. This leads to $\kappa \ll {\cal N}\sim 50$.
So, both upper limits on $\kappa$ needed for the validity of Eq.
(\ref{eq:n2}) are effectively the same. Actually, they exclude the
region only where the spectral feature itself becomes too small to
be observable.

\begin{figure}[tbh!]
\centerline{
\psfig{figure=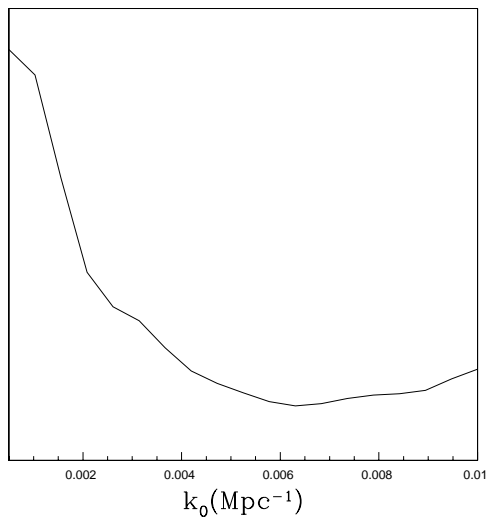,width=0.4\textwidth,angle= 0} }
\caption{\small
Marginalized posterior distributions for $k_0$.
}
\label{fig:k0}
\end{figure}

    Fig. \ref{fig:k0} gives the probability distribution
for $k_0$ (location of the feature) which is larger for smaller $k_0$
values. It is clear that the feature, if exists at all, should
preferentially lie on large scales: the marginalized upper limit for
$k_0$ is 0.00355 Mpc$^{-1}$ at 95\% CL. Therefore, our model suggests that
the main evidence for large running in
the WMAP dataset comes from sufficiently low multipoles with
$l \lesssim 40 $.$^2$ {\footnotetext[2]
{Here we use the correspondence between $k$ and $l_{eff}$ in the same
form as the WMAP team: $k R_h = l_{eff}$ with $R_h\approx 14000$ Mpc.}}

\begin{figure}[!ht]
\centerline{\psfig{figure=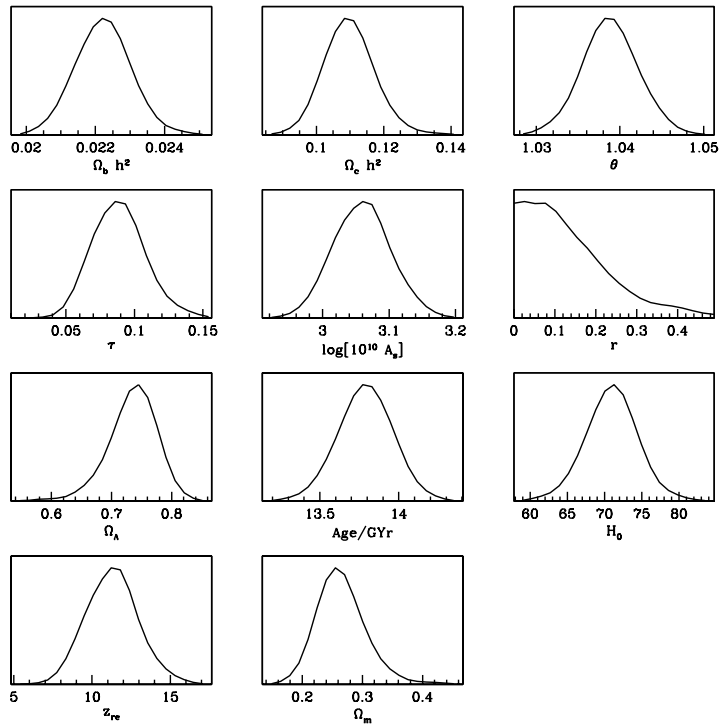,width=0.6\textwidth,angle= 0}
}
\caption{\small
Marginalized probabilities of cosmological parameters in our model.
}
\label{fig:CP}
\end{figure}

Marginalized probabilities obtained for other cosmological parameters
using the primordial spectrum given by our model
are shown in fig. \ref{fig:CP}. It follows there are no significant
changes in derived values of the cosmological parameters in comparison
with the results, obtained by the WMAP team, assuming a power-law model of
the primordial spectrum \cite{wmap5}.
\section{Conclusions and discussion}
In this paper we confront the WMAP 5-year data with
an exact solution \cite{minu} for the primordial power spectrum of
perturbations generated when the inflaton effective
potential $V(\phi)$ has a sudden small change in its second derivative
with respect to $\phi$, i.e. in $m_{\phi}^2$. The spectrum possesses a local
universal feature having the form of a step in the primordial spectral index
$n_s$, modulated by comparatively weak oscillations (by universal is meant
that the form of the spectral feature does not depend upon the structure of the
discontinuity in $V(\phi)$). It results in a large (but local) running
of $n_s$.

The simplest microscopic realization of such behaviour of
$V(\phi)$ is provided by a model having an auxiliary heavy scalar field
which experiences a rapid second order phase transition (a mini-waterfall)
during inflation in the observably accessible range of scales. Coupling
this field to the inflaton leads to the desired type of local discontinuity
in $V(\phi)$. The model is similar to that used in the hybrid inflationary
scenario, but in contrast to the latter, its parameters are chosen in such a
way that: (i) the transition occurs during inflation and not at its end,
(ii) the change in the inflaton mass is small compared to the Hubble
parameter at this moment. That is why we call it a `mini-waterfall', in contrast
to a `waterfall' which provides an end to inflation in the hybrid
scenario. Should such a mini-waterfall be detected through the corresponding
feature in the primordial power spectrum of scalar perturbations, it would
provide direct experimental evidence for the naturalness of a similar (though
larger) waterfall in the hybrid scenario (the latter not being directly observable
since it lies at a very small comoving scale).

We find that the best $\chi^2_{eff}$ for this model shows an improvement by
$3.052$ over the best fit obtained assuming a featureless power law for the
primordial spectrum. (This improvement comes at the cost of introducing two
additional parameters.) It is shown that such a feature in the primordial spectrum,
if exists at all, should lie on large scales $k_0 \lesssim 0.003$ Mpc$^{-1}$.
Anyway, this feature is not excluded by the present observational data.
Better data expected from future CMB experiments will help to settle the
question about its existence.

An interesting problem not considered in this paper is the amount of
non-Gaussianity in the statistics of primordial perturbations. Here, strictly
speaking, one needs to distinguish two different, though related cases: the
single-field inflationary model with an effective inflaton potential having the
studied type of local non-analytic behaviour, and the two-field model
(\ref{eq:hybrid}). These models produce the same results for the power
spectrum in the leading approximation, but may become non-equivalent at the
level of deviations from Gaussian behaviour. Since the second field in
the latter model is in the fast-rolling regime, one may expect larger amount
of non-Gaussianity for it. We hope to return to this question elsewhere.

\section{Acknowledgments}
We acknowledge the use of high performance computing system at IUCAA.
AAS acknowledges IUCAA hospitality as a visiting professor during the
initial stage of this project. He was also
partially supported by the grant RFBR 08-02-00923 and by the Scientific
Programme ``Elementary particles'' of the Russian Academy of Sciences.
MJ acknowledges the postdoctoral fellowship from KASI during the final
stage of this project. A. S. acknowledges BIPAC and the support of the European Research and Training Network MRTPNCT-2006 10 035863-1 (UniverseNet).

\end{document}